\documentclass[10pt]{article}

\usepackage{graphicx}
\usepackage{epstopdf}
\usepackage{titlesec}
\DeclareGraphicsExtensions{.eps}
\usepackage{float}
\usepackage{subcaption}
\usepackage{epstopdf}
\usepackage{verbatim}
\usepackage{authblk}
\usepackage{setspace}
\usepackage{color}
\usepackage{soul}
\usepackage[margin=1in]{geometry}

\makeatletter
\renewcommand{\fnum@figure}{FIG. \thefigure}
\makeatother

\title{Impact of Angular Deviation from Coincidence Site Lattice Grain Boundaries on Hydrogen Segregation and Diffusion in $\alpha$-iron}

\author[1]{Mohamed H. Hamza}
\author[2]{Mohamed A. Hendy}
\author[2,3*]{Tarek M. Hatem}
\author[1]{Jaafar A. El-Awady}
\affil[1]{Department of Mechanical Engineering, Whiting School of Engineering, Johns Hopkins University, Baltimore, MD 21218, USA}
\affil[2]{Centre for Simulation Innovation and Advanced Manufacturing, the British University in Egypt, Cairo 11837, Egypt}
\affil[3]{Faculty of Energy and Environmental Engineering, the British University in Egypt, Cairo 11837, Egypt}

\affil[*]{Corresponding author, e-mail address:  tarek.hatem@bue.edu.eg}
\date{}

\begin{document}

\maketitle

\begin{abstract}
Coincidence Site Lattice (CSL) grain boundaries (GBs) are believed to be low-energy, resistant to intergranular fracture, as well as to hydrogen embrittlement. Nevertheless, the behavior of CSL-GBs are generally confused with their angular deviations. In the current study, the effect of angular deviation from the perfect $\Sigma 3 \left(111\right)\left[1\bar{1}0\right]$ GBs in $\alpha$-iron on the hydrogen diffusion and the susceptibility of the GB to hydrogen embrittlement is investigated through molecular static and dynamics simulations. By utilizing Rice-Wang model it is shown that the ideal GB shows the highest resistance to decohesion below the hydrogen saturation limit. Finally, the hydrogen diffusivity along the ideal GB is observed to be the highest. 
\end{abstract}

\section{Introduction}

In spite of the successful development of new alloys with outstanding strength and toughness, hydrogen (H) embrittlement stays one of the most severe and most controversial types of failure that affect almost all types of metals \cite{Barnoush2010}. This can be accredited to the fact that H atoms can easily infiltrate the metal lattice. Hydrogen embrittlement typically leads to H-induced fracture even at low stress intensities \cite{mcmahon2001hydrogen}. One of the main hypotheses proposed to explain H embrittlement is the H enhanced decohesion (HEDE). In this hypothesis the increased solubility of H results in a decrease in the atom cohesive forces at the grain boundary (GB). Thus, at adequately high H concentrations the cohesive strength of the material decreases leading to an early brittle intergranular separation due to this diminishing in the cohesive forces along the GBs \cite{song2013atomic}.

Coherent twin  boundaries (CTBs) have been observed to show superior corrosion resistance in plethora of engineering alloys \cite{seita2015dual}. However, in practice, GBs are rarely ``ideal'' and their structures do not always coincide with the proper coincidence site lattice (CSL). According to the Brandon criterion, the CSL GB structure can show small angular deviation from it's ideal symmetry plane and the proximity of this deviation to a CSL structure can determine the GB traits. Additionally, this deviation is accommodated by misfit GB dislocations. Thus, such a deviation could significantly impact the segregation and diffusivity behavior of solute atoms along the GB. Indeed, it has been shown through experimental observations that many GBs in engineering materials diverge from the ideal symmetry plane \cite{herbig2014atomic,wright2002extracting}. Herbig et al. \cite{herbig2014atomic} has also shown through transmission electron microscopy and atom probe tomography study that small deviations from CSL GB orientations in ferritic steels cause significant high peaks of segregation at those GBs.

Impurity atoms and solutes tend to segregate towards GBs and other defects due to their atomic mismatch in order to reduce the free energy of the system \cite{herbig2014atomic}. Segregation energy is an important factor, particularly in determining the vulnerability of a specific GB to H embrittlement. Molecular Statics (MS) and ab initio simulations have been used broadly to calculate H segregation energies in the lattice or at different defects in metals (e.g. \cite{SONG2011,solanki2013atomistic,rajagopalan2014grain}). On the other hand, molecular dynamics (MD) studies were also conducted to predict the H-diffusivity in the lattice of $\alpha$-iron as well as on special symmetrical tilt GBs \cite{kimizuka2011effect,liu2011effects}.

Nevertheless, a comprehensive study of the impact of small angular deviation from the ideal CSL configuration on H diffusion and segregation is still missing. Consequently, the effect of the deviation angle from a perfect $\Sigma3 \left(111\right)\left[1\bar{1}0\right]$ GB in $\alpha$-iron on H segregation energy and diffusion along the GB is analyzed. First MS simulations are utilized to assess H segregation energies at the GB and free surface. Moreover, MD simulations are executed to predict the effect of the deviation angle on H diffusivity along the GB. The results are then analyzed with regard to the impact of angular deviation on the tendency for H embrittlement. The research article is organized as follows, the simulation methodology are discussed in details in Section 2. The simulation results are presented in Section 3 and discussed in Section 4. Finally, conclusions and summary are given in Section 5.

\section{Methodology}

All simulations were conducted using the large-scale atomic/molecular massively parallel simulator (LAMMPS) \cite{plimpton1995fast}, with the embedded atom method (EAM) potential for H-$\alpha$Fe developed by Ramasubramaniam et al. \cite{Ramasubramaniam2009} and subsequently modified to account for the repulsive H-H interactions by Song and Curtin \cite{song2013atomic}. The simulation cells utilized here were rectangular with a simulation cell size of $20\times20\times20$ nm$^3$ unless otherwise stated.

For bulk diffusivity simulations, periodic boundary conditions (PBCs) were employed along the $x$, $y$ and $z$ coordinates, which are parallel to the $\left[100\right]$, $\left[010\right]$, and $\left[001\right]$ crystallographic directions, respectively. Hydrogen atoms were introduced randomly at tetrahedral sites, which are the most energetically favorable sites for H occupation in BCC iron, with a concentration of $\sim 1.0\%$ of all possible occupation sites in the simulation cell.

For free surface diffusivity simulations, free surface boundary conditions were imposed along the $z$ direction, while PBCs are imposed along the other two directions. The H atoms were randomly introduced at tetrahedral sites within a 1.0 nm layer from the top and bottom free surfaces of the simulation cell with a concentration of $\sim 1.0\%$ based on all occupation sites available in those regions.

For GB diffusivity simulations, a symmetrical tilt $\Sigma{3} \left(111\right)\left[1\bar{1}0\right]$ GB with misorientation angle $\theta = 109.5^{\circ}$ was introduced at the middle of the height of the simulation cell. Periodic boundary conditions were employed along all three directions. In the GB plane this periodicity is attained by maintaining the corresponding CSL along the crystallographic orientations. In addition, atom deletion with multiple configurations was used to remove overlapping Fe atoms at the boundary to achieve an optimum GB energy and avoid the formation of residual stresses inside the simulation cell \cite{tschopp2007structures}. The H-atoms were then added at the tetrahedral sites within a $0.5$ nm thick layer surrounding the GB plane along its plane normal.

After the simulation cell was initialized, energy minimization was performed at $0$ K using the Polak-Ribiere conjugate gradient algorithm \cite{yuan2014modified} with energy tolerance of $10^{-12}$ and force tolerance of $10^{-13}$ eV/\AA. The system temperature was then increased to the desired temperature in the range of $50-700$K and then the system was relaxed for a period of $2$ ns. The time step was set to be 0.001 ps for all the segregation and diffusion simulations. The isobaric-isothermal NPT ensemble (constant number of atoms, pressure, and temperature) \cite{mendelev2003development} was used for the bulk and GB simulations, while the canonical NVT ensemble (constant number of atoms, volume, and temperature) was used for the free surfaces simulations.

As it was mentioned previously, by applying Brandon criterion for conserving CSL structures along the special GB planes, the allowable deviation angle threshold is \cite{brandon1966structure}:

\begin{equation}\label{eqn:Brandon}
\phi_{limit} = \phi_{0}\left(\Sigma\right)^{-\frac{1}{2}}
\end{equation}

\noindent where the constant $\phi_{0} \approx15^{\circ}$ and $\Sigma$ is the reciprocal value of the CSL density. Equation (\ref{eqn:Brandon}) gives $\phi_{limit} \approx 8.66^{\circ}$ for the $\Sigma{3}$ boundary. Hence, to measure the effect of a small deviation from the symmetrical tilt plane of an ideal $\Sigma{3} \left(111\right)\left[1\bar{1}0\right]$ GB on H-energetics and dynamics, the GB plane is deviated from the $\left(111\right)$ ideal symmetry plane by changing the misorientation angle $\theta$ between the two grains by a small deviation angle $0^\circ \leq \phi \leq 9.0^\circ$ measured from the $\left(111\right)$ plane, as shown in Figure \ref{Fig:Simulation_cell}. In these simulations free surface boundary conditions are employed in all three directions since the periodicity of the simulation cell can no longer be maintained. Furthermore, the same atom deletion and energy minimization technique discussed above are also utilized for these simulation cells.

In these simulations, the hydrogen GB segregation energy was computed by varying the H-atom position in the range of $0-15$ \AA ~from the GB plane and at a minimum distance of $85$ \AA ~from the simulation cell side surfaces to minimize the effect of free surface boundary conditions. Each simulation was repeated at least 5 times with different atom positions.

In addition, the H-diffusivity is also calculated at 300 K as a function of $\phi$ by randomly introducing H-atoms in the simulation cell at a minimum distance of $10$ nm from the simulation cell side surfaces such that the effects of the free surface boundary conditions are minimized. In addition, the H-atoms are initially placed within a $0.5$ nm thick layer around the GB with a concentration equal to $1\%$ of the available tetrahedral sites in that layer. Each simulation was performed with different random H-atom initialization at least 4 times. The same energy minimization techniques discussed above was also used here.

\subsection{Grain boundary energy and hydrogen segregation energy calculations}

The GB energy, $\gamma_{GB}$, is computed as follows \cite{bhattacharya2013ab}:

\begin{equation}\label{eq:GB_energy}
\gamma_{GB} = \frac{E_{GB}-E_{SC}}{A_{GB}}
\end{equation}

\noindent where $E_{GB}$ is the simulation cell total energy having both a GB and free surfaces and is calculated from the summation of the potential and kinetic energies of all atoms in the simulation cell, $E_{SC}$ is the single crystal simulation cell total energy having the same dimensions with free surfaces only, and $A_{GB}$ is the GB area.

In addition, the hydrogen GB-segregation energy, $E_{seg,GB}^{\alpha}$, is computed as follows \cite{rajagopalan2014grain,solanki2013atomistic}:

\begin{equation}\label{eq:GB_segregation}
E_{seg,GB}^{\alpha} = \left(E_{GB}^{\alpha}-E_{GB}\right)-\left(E_{SC}^{\alpha}-E_{SC}\right)
\end{equation}

\noindent where $E_{GB}^{\alpha}$ is the simulation cell total energy with a GB, free surfaces, and H-atoms, while $E_{SC}^{\alpha}$ is the corresponding single crystal simulation cell total energy with free surfaces and H-atoms.

Similarly, the hydrogen free surface segregation energy, $E_{seg,FS}^{\alpha}$, is calculated as follows \cite{hamza2015hydrogen}:

\begin{equation}\label{eq:FS_segregation}
E_{seg,FS}^{\alpha} = \left(E_{FS}^{\alpha}-E_{FS}\right)-\left(E_{SCP}^{\alpha}-E_{SCP}\right)
\end{equation}

\noindent where $E_{FS}^{\alpha}$ is the single crystal simulation cell total energy with free surface boundary conditions and charged with H-atoms, $E_{FS}$ is the corresponding pure cell total energy with free surface boundary conditions , $E_{SCP}^{\alpha}$ is the corresponding hydrogen charged cell total energy with PBCs and finally $E_{SCP}$ is the corresponding pure cell total energy with PBCs.   

\subsection{Hydrogen diffusivity calculations}

The mean square displacement (MSD) of all H-atoms during an MD simulation can be computed by \cite{liu2011effects}:

\begin{equation}\label{eq:MSD}
MSD = \frac{1}{2dN}\sum\limits_{i=1}^{N} \left(r_{i}\left(t_{0}+\Delta t\right)-r_{i}\left(t_{0}\right)\right)^2
\end{equation}

\noindent where $d$ is  a dimensionality parameter that is equal to 1.0 for diffusivity calculations along a certain direction, 2.0 for free surface and GB, and 3.0 for bulk, $N$ is the total number of H-atoms, $r_i\left(t_{0}\right)$ is the initial position of a H-atom $i$ at time $t_{0}$ and $r_i\left(t_{0}+\Delta t\right)$ is its position after a time step of $\Delta t$. The H-diffusivity is thus computed as the slope of MSD versus time. The bulk H-diffusivity is computed as the average between the diffusivity along the $x$, $y$, and $z$ directions. For free surface and GB simulations, the H-diffusivity is computed as the average of the diffusion coefficients within the free surface (i.e. $x$-$y$ plane) and the GB (i.e. $x$-$z$ plane), respectively.

\section{Results}

\subsection{Effect of small deviations from the $\Sigma3$ $\left(111\right)$ ideal symmetry plane on H-segregation energy}

\begin{figure}[p]
  \includegraphics[width=\linewidth]{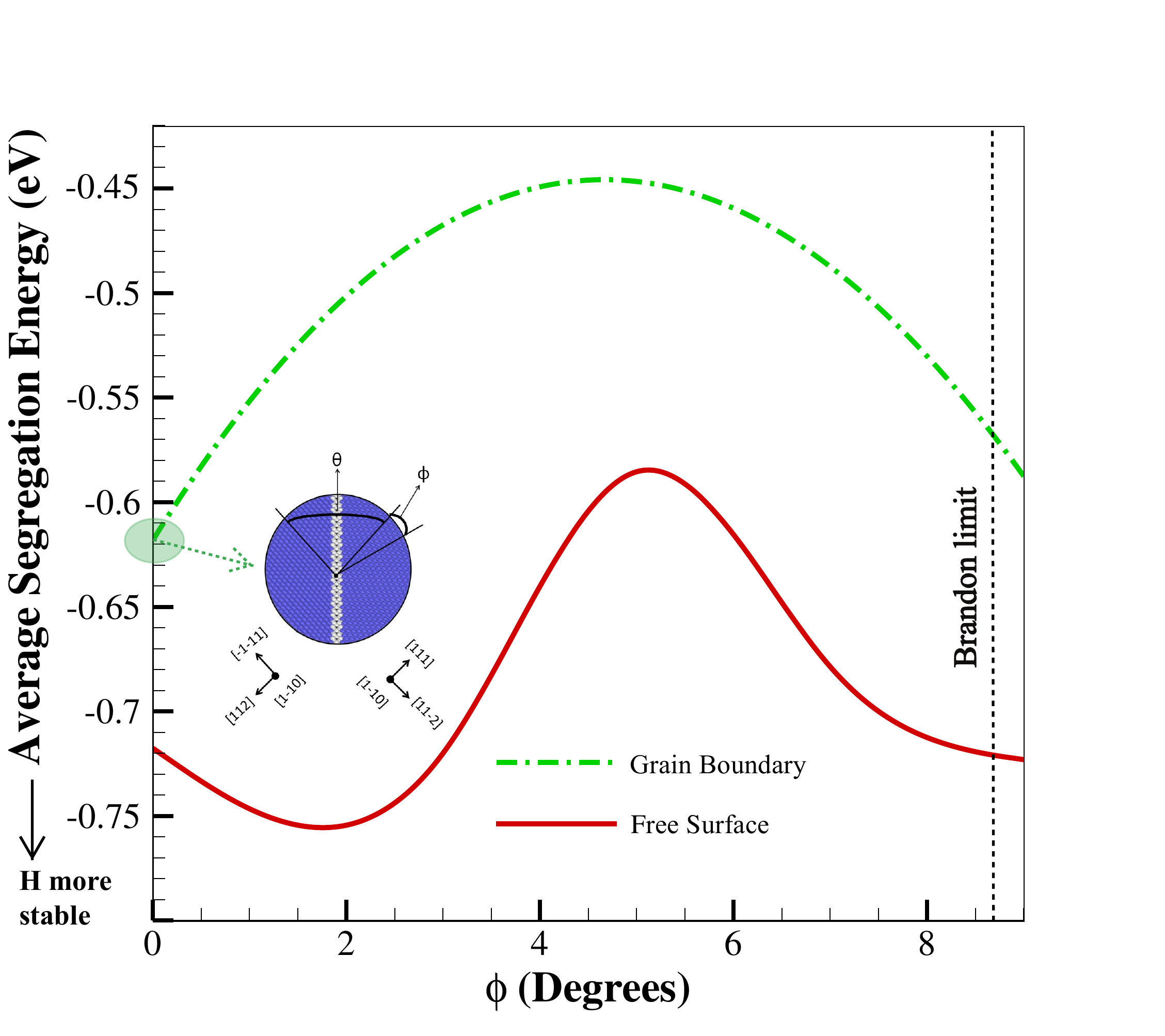}
  \caption{The average GB and free surface H-segregation energy within a 2 \AA  layer encompassing the aforementioned defects as a function of the deviation angle from the $\Sigma3$ $\left(111\right)\left[1\bar{1}0\right]$ ideal symmetry plane are presented. The dash-dot parabolic fit and the solid spline interpolation fit represents the H behavior at GB and free surface respectively (Statistical analysis can be found in the supplementary document). A Schematic diagram of the simulation cell for the $\Sigma3 \left(111\right) \left[1\bar{1}0\right]$ GB structures is embedded as well; where the misorientation angle, $\theta$, between the $(1\bar{1}\bar{1})$ plane normal and ideal $(111)$ reference plane normal as well as the deviation angle, $\phi$, are shown.}
  \label{Fig:Simulation_cell}
\end{figure}

From the current simulations the GB energy of the ideal $\Sigma3 \left(111\right) \left[1\bar{1}0\right]$ GB structures is $1.3$ J$/$m$^{2}$, which is in good agreement with previous published estimates \cite{solanki2013atomistic}. In addition, the average H-segregation energy as a function of position from the GB as computed from equation (\ref{eq:GB_segregation}) is shown in Figure \ref{Fig:Simulation_cell} (Supplementary Figure S1) for $\phi = 0^\circ, 3^\circ, 5^\circ,$ and $9^\circ$. For all the simulated cases the segregation energy at the GB is the highest, which indicates that these GBs are all more preferential for hydrogen segregation as compared to the bulk lattice. Moreover, as the angular deviations increases the GB characteristic H-absorption length increases. This length indicates the influence-zone of the GB on H-atoms located at tetrahedral sites away from the GB plane. The H-absorption length for the ideal $\Sigma{3} \left(111\right)\left[1\bar{1}0\right]$ GB is $\sim 3$ \AA, and reaches $\sim 10$ \AA ~for $\phi = 9^\circ$. This increase in the characteristic H-absorption length is mainly attributed to the increase in atomic disorder in the GB with increasing deviation angle.

The impact of the deviation angle on the H-segregation energy within a 2 \AA layer encompassing the GB is summarized in Figure \ref{Fig:Simulation_cell} (Supplementary Figure S2). The H-segregation energy is the highest in the ideal GB with a mean value of $\sim -0.62$ eV, decreases to $\sim -0.45$ eV and remains relatively constant for $3^\circ \leq \phi \leq 7^\circ$, then increases to $\sim -0.6$ eV for $\phi \leq 9^\circ$. Note all the segregation energies aforementioned are mean values calculated using statistical analysis with a $95\%$ confidence interval.

As shown in Figure \ref{Fig:Simulation_cell}, the free surface in the $\left[\bar{1}\bar{1}1\right]$ grain, termed hereafter ``Grain 1'', always has the same orientation in all simulations. However, the free surface in the other grain, termed hereafter ``Grain 2'', will change for the different cases with a deviation from the ideal GB. Thus, a different H-segregation energy would be anticipated for that later free surface. The average H-segregation energy, as computed from equation (\ref{eq:FS_segregation}), for each free surface orientation is summarized in Figure \ref{Fig:Simulation_cell} (Supplementary Figure S3). There are no considerable differences between the free surface H-segregation energy in ``Grain 1'' and ``Grain 2'' for $\phi = 0^\circ$, $3^\circ$, and $9^\circ$, in which the mean value with a $95\%$ confidence level is $\sim -0.71$ eV. On the other hand, the H-segregation energy mean value for the free surface in ``Grain 2'' with a deviation angle of $5^\circ$ is $\sim -0.44$ eV, which is the lowest value, in absolute terms, in all the simulated cases. It is also interesting to note that this energy is smaller, in absolute terms, than those for the predicted GB H-segregation energies.

\subsection{Hydrogen Diffusivity}

\begin{figure*}[p]
  \includegraphics[width=\linewidth]{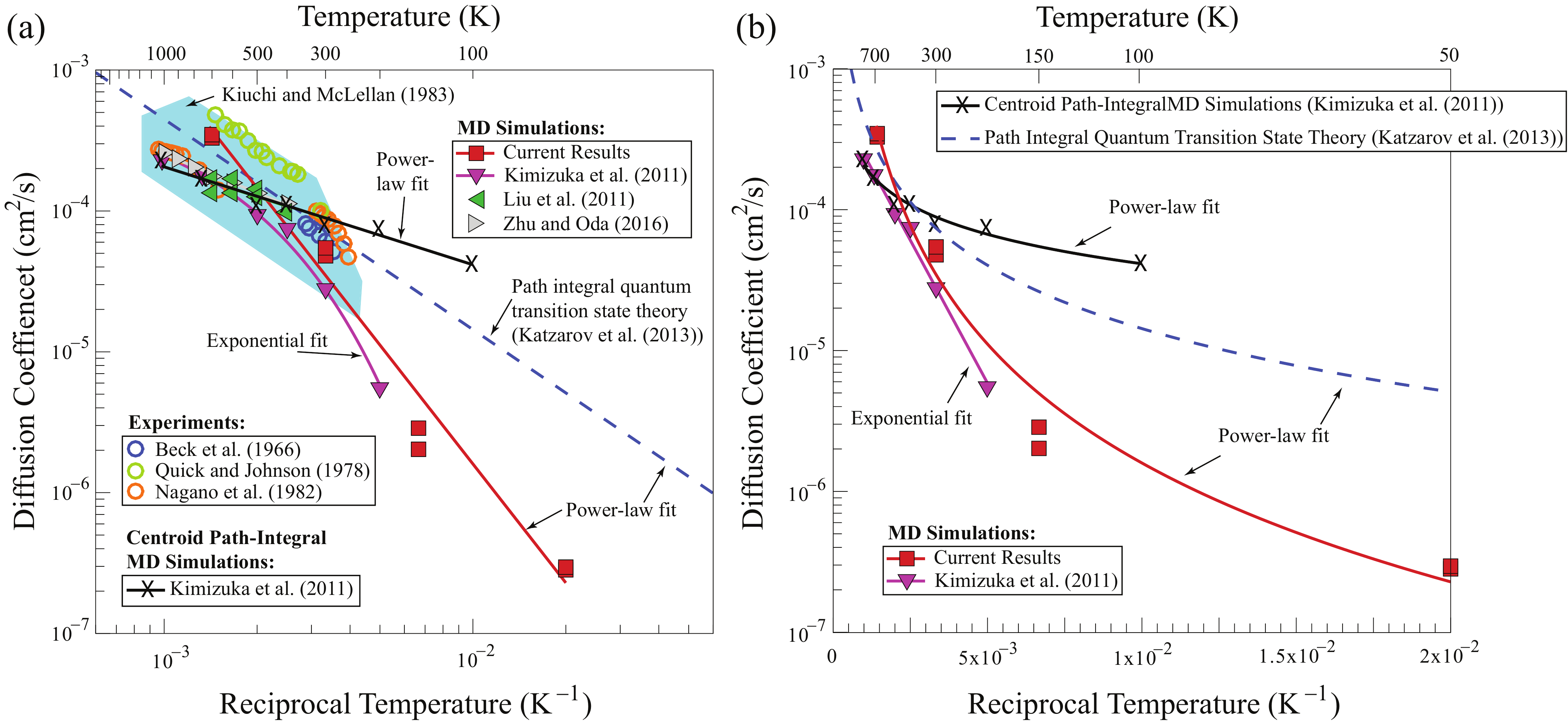}
  \caption{(a) The lattice H diffusion coefficient in $\alpha$-Fe as predicted from the current MD simulations. Other predictions from published experiments \cite{Beck1966,Quick1978,nagano1982hydrogen}, MD simulations \cite{liu2011effects,kimizuka2011effect,Zhu2016}, CMD simulations \cite{kimizuka2011effect}, and PI-QTST calculations \cite{Katzarov2013}. The shaded region represent a large rage of experimental predictions as summarized by Kiuchi and McLellan \cite{kiuchi1983solubility}. (b) A semi-log graph of a subset of data in (a) showing the non-Arrhenius relationship for the lattice H diffusion coefficient as a function of reciprocal temperature. The solid line represent the best curve fit.}
  \label{Fig:diffusion_bulk}
\end{figure*}

The predicted H lattice diffusion coefficient, $D$, from the current simulation in a bulk single crystal $\alpha$-Fe as a function of temperature is shown in Figure \ref{Fig:diffusion_bulk}a compared to previously published experimental results \cite{Beck1966,Quick1978,nagano1982hydrogen}, MD simulations \cite{liu2011effects,kimizuka2011effect,Zhu2016}, centroid molecular dynamics (CMD) simulations \cite{kimizuka2011effect}, and path integral quantum transition state theory (PI-QTST) calculations \cite{Katzarov2013}. In general, the currently predicted H diffusion coefficient as well as previously published MD simulation predictions fall in the range of experimental scatter in the temperature range of 250-1200 K. This large scatter in the experimental predictions is typically attributed to trapping of H at different defects (dislocation, grain boundaries, precipitates, etc.) among other effects \cite{kiuchi1983solubility,kimizuka2011effect}. Thus, the comparisons between the MD and experimental predictions should be made on a qualitative basis rather than a quantitative one.

In addition, as shown in Figure \ref{Fig:diffusion_bulk}b, the relationship between the diffusion coefficient and the reciprocal temperature as predicted from the current MD simulations follows a nearly non-Arrhenius relationship. There is a distinct deviation from a linear relationship between the logarithmic diffusion coefficient and the reciprocal temperatures for temperatures below 150 K. This is qualitatively in agreement with CMD \cite{kimizuka2011effect}, PI-QTST\cite{Katzarov2013}, and density function theory (DFT) \cite{DiStefano2015} predictions. This non-Arrhenius response is attributed to that H diffusion is dominated by the classical jump over the barrier mechanisms above a critical temperature, while below it quantum tunneling becomes dominant \cite{kimizuka2011effect,Katzarov2013,DiStefano2015}. Nevertheless, it should be noted that quantum tunneling effects could not be captured accurately in classical MD simulations. This explains the large deviation between the predicted diffusion coefficient below room temperature from the current MD simulations and those predicted from other methods that account for quantum tunneling. However, for $T \geq 300$ the current MD simulations are in excellent agreement with the PI-QTST and CMD predictions.

\begin{figure}[p]
  \includegraphics[width=\linewidth]{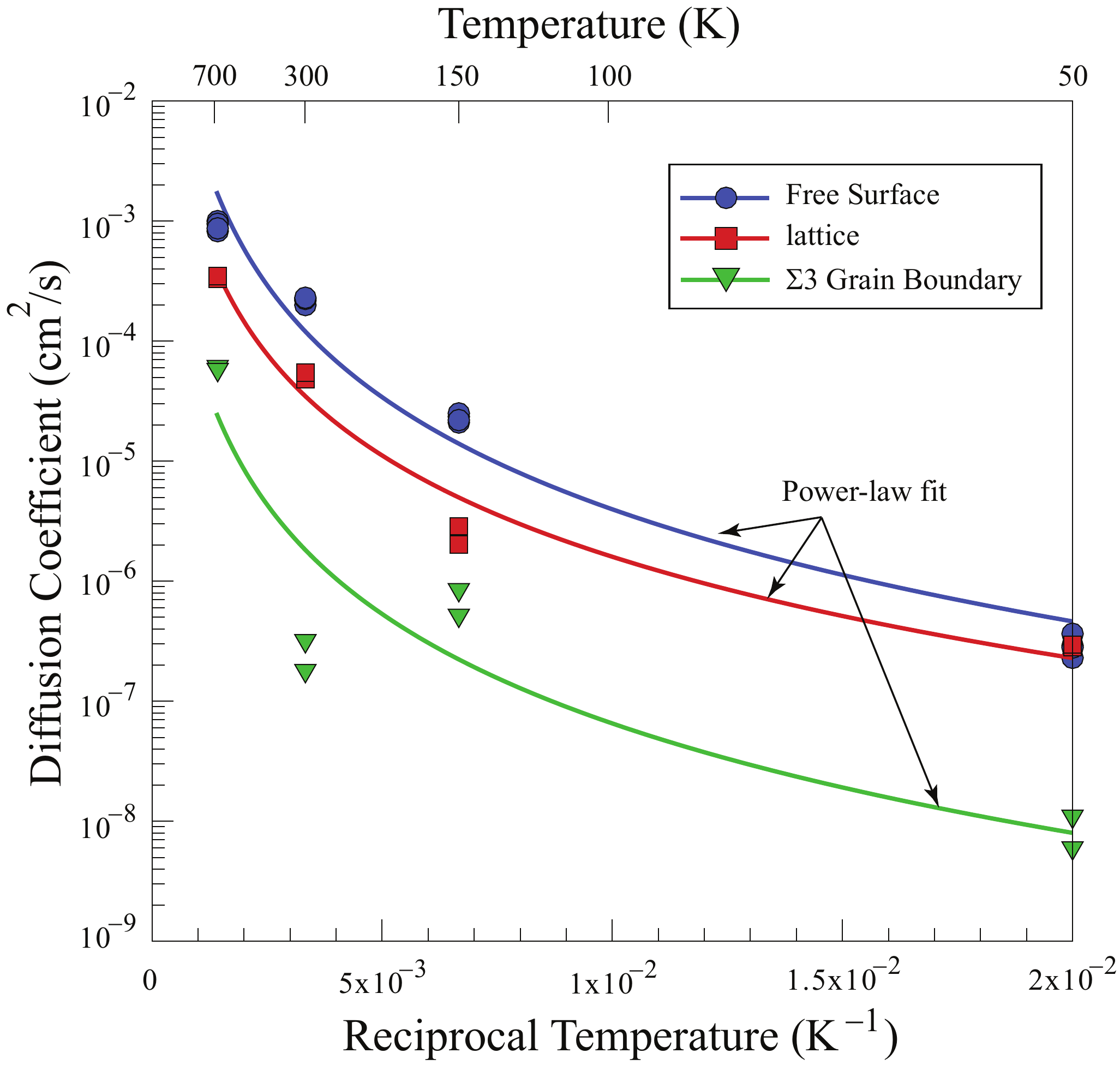}
  \caption{A semi-log graph showing the predicted H diffusion coefficient as predicted from the current MD simulations for lattice, $\left(001\right)$ free surfaces and ideal $\Sigma3 \left(111\right)\left[1\bar{1}0\right]$ GB as a function of reciprocal temperature.  The solid line represent the best power-law curve fit.}
  \label{Fig:diffusion_all}
\end{figure}

The predicted H diffusion coefficient as predicted from the current simulations for lattice, $\left(001\right)$ free surfaces and ideal $\Sigma3 \left(111\right)\left[1\bar{1}0\right]$ GB as a function of reciprocal temperature are shown on a semi-log scale in Figure \ref{Fig:diffusion_all}. It is clear that in all three cases the diffusivity follows a non-Arrhenius like behavior. Nevertheless, the results shown in Figure \ref{Fig:diffusion_all} below room temperature should only be viewed in a qualitative manner since quantum tunneling is not accounted for in the current simulations, which if accounted for would most likely lead to an order of magnitude higher diffusivity values at temperatures below room temperature.

\begin{figure}[p]
  \includegraphics[width=\linewidth]{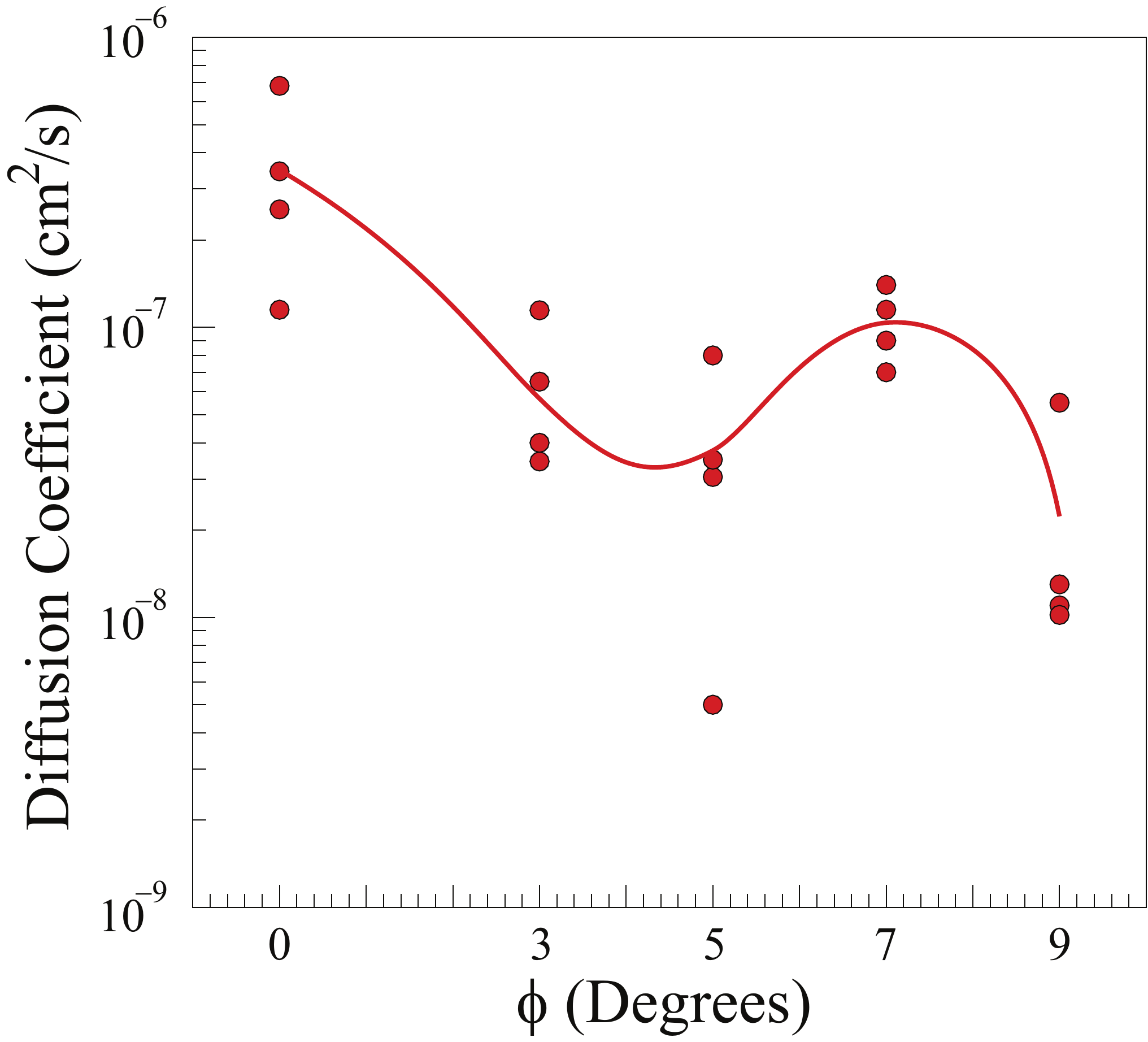}
  \caption{A semi-log graph showing the predicted H diffusion coefficient at room temperature as predicted from the current MD simulations as a function of deviation angle from the $\Sigma3$ $\left(111\right)$ ideal symmetry plane. The solid line represent the best curve fit.}
  \label{Fig:Diffusion_misorientation}
\end{figure}

The effect of the deviation from the $\Sigma3$ $\left(111\right)$ ideal symmetry plane on the GB H-diffusion coefficient at $300^\circ$ is shown in Figure \ref{Fig:Diffusion_misorientation}. The ideal GB structure is shown to result in the highest H-diffusion coefficient $D \approx 3.5\times10^{-7}$ cm$^2/$s. However, the extra atomic disorder induced by the increase in the deviation angle leads to a continuous decrease in the H-diffusivity with increasing deviation angle. In particular, the H-diffusivity for the $9^\circ$ case shows a decrease by an order of magnitude as compared to the ideal GB case. The only exception for the continuous decrease in H-diffusivity is a sudden increase for the $7^\circ$ case. This is most likely a special case in which the GB structure enhances H-diffusion as compared to the other GBs. Further quantifications of the GB structure is however needed to quantify this effect further.

\section{Discussion}

\subsection{GB Embrittlement}

Generally, solute atoms segregation at a GB can act as either GB-cohesive enhancer or GB-cohesive reducer depending on the resulting solute atom-GB interaction. As an example, Rajagopalan et al. \cite{rajagopalan2014grain} showed that phosphorous segregation at $\Sigma5 \left(210\right) \theta = 53.13^{\circ}$ GBs in $\alpha$-Fe reduces the GB-cohesion strength leading to a weakened GB structure, while vanadium acts as a GB-cohesive enhancer and strengthens this same GB. On the other hand, H typically leads to an increase in the likelihood of GB separation and leads to the formation of micro cracks at the GB, and consequently intergranular fracture in almost all metals.

In order to quantify the effect of the deviation from the $\Sigma3$ $\left(111\right)$ ideal symmetry plane on the susceptibility to H-embrittlement, the GB-cohesive energy can be computed by the Rice-Wang model \cite{rice1989embrittlement,zhong2000charge}:

\begin{equation}\label{eq:Rice}
2\gamma_{int} = \left(2\gamma_{int}\right)_{0} - \left(E_{seg,GB}^{\alpha} - E_{seg,FS}^{\alpha}\right)\tau
\end{equation}

\noindent where $2\gamma_{int}$ is the GB-cohesion energy with the existence of H atom, $\left(2\gamma_{int}\right)_0$ is the cohesion energy of the pure GB, and $\tau$ is the GB and free surface coverage constant. Based on equation (\ref{eq:Rice}), as the difference between the GB and free surface segregation energies increases, the GB-cohesion energy will be less than that of the pure GB, leading to an enhanced embrittlement effect \cite{yamaguchi2004first}.

As shown in Figure \ref{Fig:Simulation_cell}, the hydrogen free surface segregation energy is always more negative than that of the GB, and hence the GB cohesion energy as predicted from equation (\ref{eq:Rice}) will be reduced for the ideal and deviated GB cases. This implies that H is more energetically stable at the free surfaces than GB, leading to decrease in the Griffith work for brittle fracture, hence intergranular fracture is observed. Moreover, the embrittlement potency can be viewed as a linear function of the difference between the free surfaces and GB segregation energies. Accordingly, the ideal GB seems to be more resistant to H embrittlement as compared to the deviated cases \cite{yamaguchi2004first}. Compared to all the simulated cases here, the deviated GB with $\phi = 3^{\circ}$ seems to show the greatest susceptibility to decohesion and subsequently to boundary separation  as compared to the all other cases considered here.

In the case of GB saturation, where all trapping sites in the GB are occupied, the GB segregation energy will be effectively zero \cite{solanki2013atomistic}. Thus, the free surface segregation energy will be the key factor in determining the GB potency to embrittlement. As shown in Figure \ref{Fig:Simulation_cell} (Supplementary Figure S1), the $5^{\circ}$ deviation case has the lowest free surface H-segregation energy, hence this case is expected to show the highest resistance to embrittlement, followed by the $7^{\circ}$ deviation case. However, for the $0^\circ$, $3^\circ$, and $9^\circ$ cases there are no noticeable differences between their free surface segregation energies, thus they will all show similar embrittlement effects, which is relatively higher than the other GB cases simulated here.

\subsection{Effect of deviation angle on hydrogen diffusion in the grain boundary}

The deviation from the ideal GB symmetry plane will significantly affect the diffusivity of H within the GB structure. Angular deviations can be classified as either deviation from the ideal symmetry plane or deviation from the CSL misorientation angle. Brandon \cite{brandon1966structure} indicated that with a small deviation from the ideal symmetry plane, the grain boundary will be accommodated with a forest of screw dislocations superimposed on the coincidence boundary. This occurs when the sub-boundary axis of rotation is perpendicular to the CSL boundary ideal symmetry plane. On the other hand, for the case of deviation from the misorientation angle, Brandon's model suggests a superimposed boundary of mixed dislocations formed within the CSL boundary with an axis of rotation being parallel to the ideal symmetry plane \cite{brandon1966structure}. In practice, the axis of rotation of this sub-boundary structure lies at an arbitrary angle from the ideal reference plane, and thus both sub-boundary suggested models will occur simultaneously. It can be deduced from Figure \ref{Fig:Simulation_cell} that in our current simulations the angular deviation would lead to the superposition of both types of sub-boundary models.

It can thus be inferred that with the increase of dislocation densities within the GB, more trapping sites will be available \cite{krom2000hydrogen}, which subsequently impede hydrogen motion. This can also be correlated with the diffusivity trend noticed in Figure \ref{Fig:Diffusion_misorientation}. As with extra atomic mismatch and incoherence, the diffusivity along those boundaries decreases and the boundary loose its hydrogen transport ability.

\section{Conclusions}

Hydrogen segregation and diffusion along the $\Sigma3 \left(111\right)\left[1\bar{1}0\right]$ $\alpha$-iron GB structure was comprehensively analyzed using molecular static and dynamics simulations. The results strongly recommend that angular deviations from the proper coincidence site lattice significantly affects the hydrogen energetics and kinetics within the grain boundary. The simulation results indicate that the ideal GB structure acts as a preferable hydrogen sink and has the highest binding energy to the segregated atoms. However, with increasing deviation angle from the ideal symmetry plane within the limits defined by the Brandon's criterion for classification of grain boundaries it is shown that the hydrogen segregation energy at the grain boundary decreases, in absolute terms, with increases deviation angle up to a deviation angle of 5$^\circ$, and subsequently decreases. The hydrogen diffusivity along the ideal grain boundary is observed to be the highest and decreases continuously with increasing deviation angle due to the extra atomic disorder in the deviated grain boundaries. Finally, by utilizing the Rice-Wang model it is shown that the $3^{\circ}$ deviated GB has the highest susceptibility to hydrogen embrittlement, while the ideal GB shows the highest resistance to decohesion below the hydrogen saturation limit. These results suggest the importance of considering all crystallographic aspects of grain boundaries in real materials when quantifying segregation and embrittlement. The results can also help explain the experimentally observed susceptibility of coherent twin boundaries to crack initiation \cite{seita2015dual}.

\section*{Acknowledgments}

This work was supported by the National Science Foundation CAREER Award \#CMMI-1454072 and Academy of Scientific Research and Technology JESOR grant \#17.

\bibliographystyle{Hamza_MRSComm}

\providecommand{\noopsort}[1]{}\providecommand{\singleletter}[1]{#1}%


\noindent Supplementary Material Available: Statistical analysis for the grain boundary segregation energy of H solute atoms within 30 \AA \hspace{0.2em}layer are mentioned with $95\%$ confidence interval for all the tested configurations. Additionally, a summary of H behavior within 2 \AA \hspace{0.2em}layer of the grain boundaries is given with its corresponding error bars. Finally, the free surface segregation energy of H atoms for the corresponding grain boundaries types with its statistical analysis is showed. These segregation energies are thereafter used to compute the susceptibility of H embrittlement of the grain boundaries of interest as discussed in the manuscript.

\end{document}